# On classical electrodynamics in Dirac's equation matrix form

**Alexander G. Kyriakos**

*Saint-Petersburg State Institute of Technology,*
*St. Petersburg, Russia**

## Abstract

It is repeatedly marked, that Maxwell's equations can be expressed in matrix form conterminous to the form of the Dirac electron equations. But in the literature there are no consecutive and final results of this subject. In the present paper it is shown that the Maxwell theory can be finely represented in the matrix form of Dirac's equation, if the Dirac wave function is identified with the electromagnetic wave by defined way. It seems to us, that such representation allows us to see new possibilities in the connection of the classical and quantum electrodynamics.

*Present address: Athens, Greece, e-mail: agkyriak@otenet.gr

## 1. Introduction

It is noted, that the first who has paid attention to an opportunity of the matrix representation of the electrodynamics in the form of Dirac's equations was W.J.Archibald [1]. Briefly it was mentioned in the book [2]. Separate aspects of this theme were considered in several articles [3]. But full and consecutive matrix representation is absent till now. We managed to construct such representation in some specific case of electromagnetic waves.

## 2. Electrodynamics form of Dirac's equation

As it is known [4], there are two Dirac's equation forms

$$[(\hat{\alpha}_o \hat{\varepsilon} + c\hat{\vec{\alpha}} \ \hat{\vec{p}}) + \hat{\beta} \ mc^2]\psi = 0 , \tag{2.1}$$

$$\psi^+ [(\hat{\alpha}_o \hat{\varepsilon} - c\hat{\vec{\alpha}} \ \hat{\vec{p}}) - \hat{\beta} \ mc^2] = 0 , \tag{2.2}$$

which correspond to two signs of the relativistic energy expression:

$$\varepsilon = \pm\sqrt{c^2 \vec{p}^2 + m^2 c^4} , \tag{2.3}$$

Here $\hat{\varepsilon} = i\hbar \dfrac{\partial}{\partial t}$, $\hat{\vec{p}} = -i\hbar \vec{\nabla}$ are the operators of energy and momentum, $\varepsilon$, $\vec{p}$ are the electron energy and momentum, $c$ is the light velocity, $m$ is the electron mass, $\psi$ is the wave function ($\psi^+$ is the Hermithian-conjugate wave function) named bispinor and $\hat{\alpha}_o = \hat{1}$, $\hat{\vec{\alpha}}$, $\hat{\alpha}_4 \equiv \hat{\beta}$ are the Dirac matrices:

$$\hat{\alpha}_1 = \begin{pmatrix} 0 & 0 & 0 & 1 \\ 0 & 0 & 1 & 0 \\ 0 & 1 & 0 & 0 \\ 1 & 0 & 0 & 0 \end{pmatrix}, \ \hat{\alpha}_2 = \begin{pmatrix} 0 & 0 & 0 & -i \\ 0 & 0 & i & 0 \\ 0 & -i & 0 & 0 \\ i & 0 & 0 & 0 \end{pmatrix}, \ \hat{\alpha}_3 = \begin{pmatrix} 0 & 0 & 1 & 0 \\ 0 & 0 & 0 & -1 \\ 1 & 0 & 0 & 0 \\ 0 & -1 & 0 & 0 \end{pmatrix}, \ \vec{\alpha}_4 = \begin{pmatrix} 1 & 0 & 0 & 0 \\ 0 & 1 & 0 & 0 \\ 0 & 0 & -1 & 0 \\ 0 & 0 & 0 & -1 \end{pmatrix} \tag{2.4}$$

Also as it is known, for each sign of equation (2.3) there are two Hermithian-conjugate Dirac's equations. Here we consider the electrodynamics meaning of all these equations.

Let us consider at first two Hermithian-conjugate equations, corresponding to the minus sign in (2.3):

$$\left[\left(\hat{\alpha}_o \hat{\varepsilon} + c\hat{\vec{\alpha}}\ \hat{\vec{p}}\right) + \hat{\beta}\ mc^2\right]\psi = 0, \qquad (2.5')$$

$$\psi^+\left[\left(\hat{\alpha}_o \hat{\varepsilon} + c\hat{\vec{\alpha}}\ \hat{\vec{p}}\right) + \hat{\beta}\ mc^2\right] = 0, \qquad (2.5'')$$

We will consider the wave function $\psi$ as electromagnetic wave function. Put, e.g., $\psi = \psi(y)$ and choose

$$\psi = \begin{pmatrix}\psi_1 \\ \psi_2 \\ \psi_3 \\ \psi_4\end{pmatrix} = \begin{pmatrix}E_x \\ E_z \\ iH_x \\ iH_z\end{pmatrix}, \qquad (2.6)$$

Then
$$\psi^+ = \begin{pmatrix}E_x & E_z & -iH_x & -iH_z\end{pmatrix}, \qquad (2.7)$$

Using (2.6) and (2.7) from (2.5') and (2.5'') we obtain:

$$\begin{cases}\dfrac{1}{c}\dfrac{\partial E_x}{\partial t} - \dfrac{\partial H_z}{\partial y} + i\dfrac{\omega}{c}E_x = 0, \\ \dfrac{1}{c}\dfrac{\partial E_z}{\partial t} + \dfrac{\partial H_x}{\partial y} + i\dfrac{\omega}{c}E_z = 0, \\ \dfrac{1}{c}\dfrac{\partial H_x}{\partial t} + \dfrac{\partial E_z}{\partial y} - i\dfrac{\omega}{c}H_x = 0, \\ \dfrac{1}{c}\dfrac{\partial H_z}{\partial t} - \dfrac{\partial E_x}{\partial y} - i\dfrac{\omega}{c}H_z = 0,\end{cases} (2.8), \quad \begin{cases}\dfrac{1}{c}\dfrac{\partial E_x}{\partial t} - \dfrac{\partial H_z}{\partial y} - i\dfrac{\omega}{c}E_x = 0, \\ \dfrac{1}{c}\dfrac{\partial E_z}{\partial t} + \dfrac{\partial H_x}{\partial y} - i\dfrac{\omega}{c}E_z = 0, \\ \dfrac{1}{c}\dfrac{\partial H_x}{\partial t} + \dfrac{\partial E_z}{\partial y} + i\dfrac{\omega}{c}H_x = 0, \\ \dfrac{1}{c}\dfrac{\partial H_z}{\partial t} - \dfrac{\partial E_x}{\partial y} + i\dfrac{\omega}{c}H_z = 0,\end{cases} (2.9)$$

where $\omega = \dfrac{mc^2}{\hbar}$. The equations (2.8, 2.9) are the Maxwell equations with complex currents [5]. (It is interesting that together with the electrical current the magnetic current also exists here. This current is equal to zero by Maxwell's theory, but its existence by Dirac doesn't contradict to the quantum theory).

As we see the equations (2.8) and (2.9) are differed by the current directions. We could foresee this result before the calculations, since the functions $\psi^+$ and $\psi$ are differed by the argument signs:

$$\psi^+ = \psi_0 e^{-i\omega t} \text{ and } \psi = \psi_0 e^{i\omega t}.$$

Let us compare now the equations corresponding to the both plus and minus signs of (2.3). For the plus sign of (2.2) we have following two equations:

$$[(\hat{\alpha}_o \hat{\varepsilon} - c\hat{\vec{\alpha}} \; \hat{\vec{p}}) - \hat{\beta} \; mc^2] \psi = 0 , \qquad (2.10)$$

$$\psi^+ [(\hat{\alpha}_o \hat{\varepsilon} - c\hat{\vec{\alpha}} \; \hat{\vec{p}}) - \hat{\beta} \; mc^2] = 0 , \qquad (2.11)$$

The electromagnetic form of the equation (2.10) is:

$$\begin{cases} \dfrac{1}{c}\dfrac{\partial E_x}{\partial t} + \dfrac{\partial H_z}{\partial y} + i\dfrac{\omega}{c} E_x = 0, \\ \dfrac{1}{c}\dfrac{\partial E_z}{\partial t} - \dfrac{\partial H_x}{\partial y} + i\dfrac{\omega}{c} E_z = 0, \\ \dfrac{1}{c}\dfrac{\partial H_x}{\partial t} - \dfrac{\partial E_z}{\partial y} - i\dfrac{\omega}{c} H_x = 0, \\ \dfrac{1}{c}\dfrac{\partial H_z}{\partial t} + \dfrac{\partial E_x}{\partial y} - i\dfrac{\omega}{c} H_z = 0, \end{cases} \qquad (2.12)$$

Obviously, the electromagnetic form of equation (2.11) will have the opposite signs of the currents compared to (2.12).

Comparing (2.12) and (2.8) we see that the equation (2.12) can be considered as Maxwell's equation of the retarded wave. So as not to use the retarded wave, together with the wave function of the advanced wave $\psi_{adv}$ we can consider the wave function of retarded wave in the form:

$$\psi_{ret} = \begin{pmatrix} E_x \\ -E_z \\ iH_x \\ -iH_z \end{pmatrix} , \qquad (2.13)$$

Then, contrary to the system (2.12) we get the system (2.9). The transformation of the function $\psi_{ret}$ to the function $\psi_{adv}$ is named in the quantum mechanics the charge conjugation operation.

## 2.1. Electrodynamics' sense of bilinear forms

Enumerate the main Dirac's matrices [4]:

1) $\hat{\alpha}_4 \equiv \hat{\beta}$ is the scalar,   2) $\hat{\alpha}_\mu = \{\hat{\alpha}_0, \hat{\vec{\alpha}}\} \equiv \{\hat{\alpha}_0, \hat{\alpha}_1, \hat{\alpha}_2, \hat{\alpha}_3\}$ is the 4-vector,

3) $\hat{\alpha}_5 = \hat{\alpha}_1 \cdot \hat{\alpha}_2 \cdot \hat{\alpha}_3 \cdot \hat{\alpha}_4$ is the pseudoscalar.

Using (2.6-2.7) and taking in account that $\psi = \psi(y)$ it is easy to obtain the electrodynamics' expression of bispinors, corresponding to these matrices:

1) $\psi^+ \hat{\alpha}_4 \psi = (E_x^2 + E_z^2) - (H_x^2 + H_z^2) = \vec{E}^2 - \vec{H}^2 = 8\pi I_1$, where $I_1$ is the first scalar of Maxwell's theory;

2) $\psi^+ \hat{\alpha}_o \psi = \vec{E}^2 + \vec{H}^2 = 8\pi U$ and $\psi^+ \hat{\alpha}_y \psi = 8\pi c g_y$. Thus, the electrodynamics form of the 4-vector bispinor value is the energy-momentum 4-vector $\left\{ \frac{1}{c} U, \vec{g} \right\}$.

3) $\psi^+ \hat{\alpha}_5 \psi = 2(E_x H_x + E_z H_z) = 2(\vec{E} \cdot \vec{H})$ is the pseudoscalar and $(\vec{E} \cdot \vec{H})^2 = I_2$ is the second scalar of the electromagnetic field theory.

From Dirac's equation the probability continuity equation can be obtained:

$$\frac{\partial P_{pr}(\vec{r},t)}{\partial t} + div\, \vec{S}_{pr}(\vec{r},t) = 0, \qquad (2.14)$$

Here $P_{pr}(\vec{r},t) = \psi^+ \hat{\alpha}_0 \psi$ is the probability density, and $\vec{S}_{pr}(\vec{r},t) = -c\psi^+ \hat{\vec{\alpha}} \psi$ is the probability-flux density. Using the above results we can obtain: $P_{pr}(\vec{r},t) = 8\pi U$ and $\vec{S}_{pr} = c^2 \vec{g} = 8\pi \vec{S}$. Then the equation (2.14) has the view:

$$\frac{\partial U}{\partial t} + div\, \vec{S} = 0, \qquad (2.15)$$

which is the energy conservation law of the electron electromagnetic field.

## 3. Electrodynamics sense of the matrix choice

As we saw above, the matrix sequence $(\hat{\alpha}_1, \hat{\alpha}_2, \hat{\alpha}_3)$ agrees to the electromagnetic wave having $-y$-direction. But herewith only the $\hat{\alpha}_2$-matrix is "working", and other two matrices don't give the terms in the equation. The verification of this fact is the Poynting vector calculation: the bilinear forms of $\hat{\alpha}_1, \hat{\alpha}_3$-matrices are equal to zero, and only the matrix $\hat{\alpha}_2$ gives the right non-zero component of Poynting's vector.

The question arises how to describe the waves having $x$ and $z$-directions. It is not difficult to see that the matrices' sequence is not determined by the some special requirements. In fact, this matrices' sequence can be changed without breaking some quantum electrodynamics results [4, 6].

So we can write three groups of matrices, each of which corresponds to the one and only one direction, introducing the axes' indexes, which indicate the electromagnetic wave direction:

$$(\hat{\alpha}_{1x}, \hat{\alpha}_{2y}, \alpha_{3z}), \quad (\hat{\alpha}_{2x}, \hat{\alpha}_{3y}, \hat{\alpha}_{1z}), \quad (\hat{\alpha}_{2z}, \hat{\alpha}_{1y}, \hat{\alpha}_{3x}).$$

Let us choose now the wave function forms, which give the correct Maxwell equations. We will take as initial the form for the $-y$ - direction, which we already used, and from them, by means of the indexes' transposition around the circle we will get other forms for $x$ and $y$ - directions.

Since in this case the Poynting vector has the minus sign, we can suppose that the transposition takes place counterclockwise. Thus, let us check the Poynting vector values:

1) $\psi = \psi(y)$, $(\hat{\alpha}_{1x}, \hat{\alpha}_{2y}, \alpha_{3z})$, $\psi = \begin{pmatrix} E_x \\ E_z \\ iH_x \\ iH_z \end{pmatrix}$, $\psi^+ = \begin{pmatrix} E_x & E_z & -iH_x & -iH_z \end{pmatrix}$; (3.1)

$$\psi^+ \hat{\alpha}_{1x} \psi = \begin{pmatrix} E_x & E_z & -iH_x & -iH_z \end{pmatrix} \begin{pmatrix} iH_z \\ iH_x \\ E_z \\ E_x \end{pmatrix} = iE_x H_z + iE_z H_x - iE_z H_x - iE_x H_z = 0,$$

$$\psi^+ \hat{\alpha}_{2y} \psi = -(E_z H_x - E_x H_z) = -2[\vec{E} \times \vec{H}]_y, \quad \psi^+ \hat{\alpha}_{3z} \psi = 0.$$

2) $\psi = \psi(x)$, $(\hat{\alpha}_{2x}, \hat{\alpha}_{3y}, \hat{\alpha}_{1z},)$, $\psi = \begin{pmatrix} E_z \\ E_y \\ iH_z \\ iH_y \end{pmatrix}$, $\psi^+ = \begin{pmatrix} E_z & E_y & -iH_z & -iH_y \end{pmatrix}$; (3.2)

$$\psi^+ \hat{\alpha}_{2x} \psi = -2(E_y H_z - E_z H_y) = -2[\vec{E} \times \vec{H}]_x, \quad \psi^+ \hat{\alpha}_{3y} \psi = 0, \quad \psi^+ \hat{\alpha}_{1z} \psi = 0.$$

3) $\psi = \psi(z)$, $(\hat{\alpha}_{2z}, \hat{\alpha}_{1y}, \hat{\alpha}_{3x})$, $\psi = \begin{pmatrix} E_y \\ E_x \\ iH_y \\ iH_x \end{pmatrix}$, $\psi^+ = \begin{pmatrix} E_y & E_x & -iH_y & -iH_x \end{pmatrix}$; (3.3)

$$\psi^+ \hat{\alpha}_{3x} \psi = 0, \quad \psi^+ \hat{\alpha}_{1y} \psi = 0, \quad \psi^+ \hat{\alpha}_{2z} \psi = -2(E_x H_y - E_y H_x) = -2[\vec{E} \times \vec{H}]_z.$$

As we see, we took the correct result: by the counterclockwise indexes' transposition the wave functions describe the electromagnetic wave, which are moved in negative directions of the corresponding co-ordinate axes.

We may hope that by the clockwise indexes' transposition, the wave functions will describe the electromagnetic wave, which are moved in positive directions of co-ordinate axes. Prove this:

1) $\psi = \psi(y)$, $(\hat{\alpha}_{1_x}, \hat{\alpha}_{2_y}, \alpha_{3_z})$, $\psi = \begin{pmatrix} E_z \\ E_x \\ iH_z \\ iH_x \end{pmatrix}$, $\psi^+ = (E_z \ E_x \ -iH_z \ -iH_x)$; (3.4)

$\psi^+ \hat{\alpha}_{1_x} \psi = 0$, $\psi^+ \hat{\alpha}_{2_y} \psi = 2(E_z H_x - E_x H_z) = 2[\vec{E} \times \vec{H}]_y$, $\psi^+ \hat{\alpha}_{3_z} \psi = 0$,

2) $\psi = \psi(x)$, $(\hat{\alpha}_{2_x}, \hat{\alpha}_{3_y}, \hat{\alpha}_{1_z},)$, $\psi = \begin{pmatrix} E_y \\ E_z \\ iH_y \\ iH_z \end{pmatrix}$, $\psi^+ = (E_y \ E_z \ -iH_y \ -iH_z)$; (3.5)

$\psi^+ \hat{\alpha}_{2_x} \psi = 2(E_y H_z - E_z H_y) = 2[\vec{E} \times \vec{H}]_x$, $\psi^+ \hat{\alpha}_{3_y} \psi = 0$, $\psi^+ \hat{\alpha}_{1_z} \psi = 0$.

3) $\psi = \psi(z)$, $(\hat{\alpha}_{2_z}, \hat{\alpha}_{1_y}, \hat{\alpha}_{3_x})$, $\psi = \begin{pmatrix} E_x \\ E_y \\ iH_x \\ iH_y \end{pmatrix}$, $\psi^+ = (E_x \ E_y \ -iH_x \ -iH_y)$; (3.6)

$\psi^+ \hat{\alpha}_{3_x} \psi = 0$, $\psi^+ \hat{\alpha}_{1_y} \psi = 0$, $\psi^+ \hat{\alpha}_{2_z} \psi = 2(E_x H_y - E_y H_x) = 2[\vec{E} \times \vec{H}]_z$.

As we see, we also get the correct results.

Now we will prove that the above choice of the matrices give the correct electromagnetic equation forms. Using for example the bispinor Dirac's equation (2.10) and transposing the indexes clockwise we obtain for the positive direction of the electromagnetic wave the following results for $x$, $y$, $z$-directions correspondingly:

$$\begin{cases} \dfrac{1}{c}\dfrac{\partial E_y}{\partial t} + \left(\dfrac{\partial H_z}{\partial x}\right) = -i\dfrac{\omega}{c}E_y, \\ \dfrac{1}{c}\dfrac{\partial E_z}{\partial t} - \left(\dfrac{\partial H_y}{\partial x}\right) = -i\dfrac{\omega}{c}E_z, \\ \dfrac{1}{c}\dfrac{\partial H_y}{\partial t} - \left(\dfrac{\partial E_z}{\partial x}\right) = i\dfrac{\omega}{c}H_y, \\ \dfrac{1}{c}\dfrac{\partial H_z}{\partial t} + \left(\dfrac{\partial E_y}{\partial x}\right) = i\dfrac{\omega}{c}H_z, \end{cases} \begin{cases} \dfrac{1}{c}\dfrac{\partial E_z}{\partial t} + \left(\dfrac{\partial H_x}{\partial y}\right) = -i\dfrac{\omega}{c}E_z, \\ \dfrac{1}{c}\dfrac{\partial E_x}{\partial t} - \left(\dfrac{\partial H_z}{\partial y}\right) = -i\dfrac{\omega}{c}E_x, \\ \dfrac{1}{c}\dfrac{\partial H_z}{\partial t} - \left(\dfrac{\partial E_x}{\partial y}\right) = i\dfrac{\omega}{c}H_z, \\ \dfrac{1}{c}\dfrac{\partial H_x}{\partial t} + \left(\dfrac{\partial E_z}{\partial y}\right) = i\dfrac{\omega}{c}H_x, \end{cases} \begin{cases} \dfrac{1}{c}\dfrac{\partial E_x}{\partial t} + \left(\dfrac{\partial H_y}{\partial z}\right) = -i\dfrac{\omega}{c}E_x, \\ \dfrac{1}{c}\dfrac{\partial E_y}{\partial t} - \left(\dfrac{\partial H_x}{\partial z}\right) = -i\dfrac{\omega}{c}E_y, \\ \dfrac{1}{c}\dfrac{\partial H_x}{\partial t} - \left(\dfrac{\partial E_y}{\partial z}\right) = i\dfrac{\omega}{c}H_x, \\ \dfrac{1}{c}\dfrac{\partial H_y}{\partial t} + \left(\dfrac{\partial E_x}{\partial z}\right) = i\dfrac{\omega}{c}H_y, \end{cases}$$ (3.7)

So we have obtained three equation groups, each of which contains four equations, as it is necessary for the description of all electromagnetic wave directions. In the same way for all the other Dirac's equation forms the analogue results can be obtained.

Obviously, it is possible via canonical transformations to choose the Dirac matrices so that the electromagnetic wave could have any direction.

### 3.1. Electromagnetic form of canonical transformations of Dirac's matrixes and bispinors

As it is known [2,7], the transition from some independent variables to others, made by means of the unitary operator, is called canonical transformation.

Let us consider the electromagnetic form of the canonical transformations of matrixes and the wave functions of Dirac's equation. The choice of matrixes $\alpha$, made by us (2.4), is not unique. In our case there is a free transformation of a kind:

$$\alpha = S\, a'\, S', \qquad (3.8)$$

where $S$ is a unitary matrix, which consists of four lines and four columns. The following substitution in regard to functions $\psi'$ corresponds to this transformation:

$$\psi = S\, \psi', \qquad (3.9)$$

If we choose matrixes $\alpha'$ as:

$$\hat{\alpha}'_1 = \begin{pmatrix} 0 & 1 & 0 & 0 \\ 1 & 0 & 0 & 0 \\ 0 & 0 & 0 & 1 \\ 0 & 0 & 1 & 0 \end{pmatrix},\quad \hat{\alpha}'_2 = \begin{pmatrix} 0 & -i & 0 & 0 \\ i & 0 & 0 & 0 \\ 0 & 0 & 0 & i \\ 0 & 0 & i & 0 \end{pmatrix},\quad \hat{\alpha}'_3 = \begin{pmatrix} 1 & 0 & 0 & 0 \\ 0 & -1 & 0 & 0 \\ 0 & 0 & 1 & 0 \\ 0 & 0 & 0 & -1 \end{pmatrix};$$

$$\vec{\alpha}'_4 = \begin{pmatrix} 0 & 0 & 0 & -1 \\ 0 & 0 & 1 & 0 \\ 0 & 1 & 0 & 0 \\ -1 & 0 & 0 & 0 \end{pmatrix},\quad \hat{\alpha}'_5 = \begin{pmatrix} 0 & 0 & 0 & -i \\ 0 & 0 & i & 0 \\ 0 & -i & 0 & 0 \\ i & 0 & 0 & 0 \end{pmatrix}; \qquad (3.10)$$

then the functions $\psi$ will be connected to functions $\psi'$ according to the relationship:

$$\psi = \frac{\psi'_1 - \psi'_4}{\sqrt{2}}, \quad \psi = \frac{\psi'_2 + \psi'_3}{\sqrt{2}}, \quad \psi = \frac{\psi'_1 + \psi'_4}{\sqrt{2}}, \quad \psi = \frac{\psi'_2 - \psi'_3}{\sqrt{2}}, \quad (3,11)$$

The unitary matrix $S$, appropriate to this transformation, is equal to:

$$S = \frac{1}{\sqrt{2}} \begin{bmatrix} 1 & 0 & 0 & -1 \\ 0 & 1 & 1 & 0 \\ 1 & 0 & 0 & 1 \\ 0 & 1 & -1 & 0 \end{bmatrix}, \quad (3.12)$$

It is not difficult to check, that from this transformation we also will receive the electromagnetic forms of Maxwell's theory. Actually, for example, for the chosen form of function (2.4), using the above-stated transformations, it is easy to receive:

$$\frac{\psi'_1 - \psi'_4}{\sqrt{2}} = E_x, \quad \frac{\psi'_2 + \psi'_3}{\sqrt{2}} = E_z, \quad \frac{\psi'_1 + \psi'_4}{\sqrt{2}} = iH_x, \quad \frac{\psi'_2 - \psi'_3}{\sqrt{2}} = iH_z, \quad (3.13)$$

whence:

$$\psi' = \frac{\sqrt{2}}{2} \begin{pmatrix} (E_x + iH_x) \\ (E_z + iH_z) \\ (E_z - iH_z) \\ -(E_x - iH_x) \end{pmatrix}, \quad (3.14)$$

It is easy to be convinced, substituting these functions in Dirac's equation, that we will receive Maxwell's equations (in double quantity).

It is possible to assume, that the functions $\psi'$ correspond to the electromagnetic wave, moving under the angle of 45 degrees to both coordinate axes.

## 3.2. Matrix form of electromagnetic wave equation

Using (2.6, 2.7), we can write the equation of the electromagnetic wave moved along the any axis in form:

$$\left( \hat{\varepsilon}^2 - c^2 \hat{p}^2 \right) \psi = 0, \quad (3.15)$$

The equation (3.15) can also be written in the following form:

$$\left[ \left( \hat{\alpha}_o \hat{\varepsilon} \right)^2 - c^2 \left( \hat{\vec{\alpha}} \, \hat{\vec{p}} \right)^2 \right] \psi = 0, \quad (3.16)$$

In fact, taking into account that

$$\left(\hat{\alpha}_o\hat{\varepsilon}\right)^2 = \hat{\varepsilon}^2, \quad \left(\hat{\vec{\alpha}}\ \hat{\vec{p}}\right)^2 = \hat{\vec{p}}^2,$$

we see that equations (3.15) and (3.16) are equivalent.

Factorizing (3.16) and multiplying it from left on Hermithian-conjugate function $\psi^+$ we get:

$$\psi^+\left(\hat{\alpha}_o\hat{\varepsilon} - c\hat{\vec{\alpha}}\ \hat{\vec{p}}\right)\left(\hat{\alpha}_o\hat{\varepsilon} + c\hat{\vec{\alpha}}\ \hat{\vec{p}}\right)\psi = 0, \qquad (3.17)$$

The equation (3.17) may be disintegrated on two equations:

$$\psi^+\left(\hat{\alpha}_o\hat{\varepsilon} - c\hat{\vec{\alpha}}\ \hat{\vec{p}}\right) = 0, \qquad (3.18)$$

$$\left(\hat{\alpha}_o\hat{\varepsilon} + c\hat{\vec{\alpha}}\ \hat{\vec{p}}\right)\psi = 0, \qquad (3.19)$$

It is not difficult to show (using (2.6, 2.7) that the equations (3.18) and (3.19) are Maxwell's equations without current, or in other words, they are the Maxwell equation for the electromagnetic wave.

### 3.3. The electromagnetic form of electron theory Lagrangian

As it is known [5], the Lagrangian of the free field Maxwell's theory is:

$$L_M = \frac{1}{8\pi}\left(\vec{E}^2 - \vec{H}^2\right), \qquad (3.20)$$

and as Lagrangian of Dirac's theory can be taken the expression [4]:

$$L_D = \psi^+\left(\hat{\varepsilon} + c\hat{\vec{\alpha}}\ \hat{\vec{p}} + \hat{\beta}\ mc^2\right)\psi, \qquad (3.21)$$

For the electromagnetic wave moving along the $-y$-axis the equation (3.21) can be written:

$$L_D = \frac{1}{c}\psi^+\frac{\partial\psi}{\partial t} - \psi^+\hat{\alpha}_y\frac{\partial\psi}{\partial y} - i\frac{mc}{\hbar}\psi^+\hat{\beta}\ \psi, \qquad (3.22)$$

Transferring the each term of (3.22) in electrodynamics' form we obtain for the electromagnetic wave equation Lagrangian:

$$L_s = \frac{\partial U}{\partial t} + div\ \vec{S} - i\frac{\omega_s}{8\pi}\left(\vec{E}^2 - \vec{H}^2\right), \qquad (3.23)$$

where $\omega_s = \dfrac{2mc^2}{\hbar}$ (note that we must differ the complex conjugate field vectors $\vec{E}^*, \vec{H}^*$ and $\vec{E}, \vec{H}$).

The equation (3.23) can be written in other form. Using electrical and "magnetic" currents $j_\tau^e = i\frac{\omega_s}{4\pi}\vec{E}$ and $j_\tau^m = i\frac{\omega_s}{4\pi}\vec{H}$ we take:

$$L_s = \frac{\partial U}{\partial t} + div\ \vec{S} - \frac{1}{2}\left(\vec{j}_\tau^e \vec{E} - \vec{j}_\tau^m \vec{H}\right) \quad , \tag{3.24}$$

Since $L_s = 0$ thanks to (2.1), we take the equation:

$$\frac{\partial U}{\partial t} + div\ \vec{S} - \frac{1}{2}\left(\vec{j}_\tau^e \vec{E} - \vec{j}_\tau^m \vec{H}\right) = 0 , \tag{3.25}$$

which is the energy-momentum conservation law for the Maxwell equation with current.

## 4. Non-linear electromagnetic equation of electron

The stability of electron is possible only by the electrons parts self-action. We introduce the self-field interaction to the electron equation like the external field is introduced to the quantum [4] and classical [5] field equations (putting the electron mass equal to zero). Then we obtain non-linear integral-differential equation:

$$\left[\hat{\alpha}_0\left(\hat{\varepsilon} - \varepsilon_s\right) + c\hat{\vec{\alpha}}\left(\hat{\vec{p}} - \vec{p}_s\right)\right]\psi = 0 , \tag{4.1}$$

In the electromagnetic form we have:

$$\varepsilon_s = \int_0^\infty U\ d\tau = \frac{1}{8\pi}\int_0^\infty\left(\vec{E}^2 + \vec{H}^2\right)d\tau , \tag{4.2}$$

$$\vec{p}_s = \int_0^\infty \vec{g}\ d\tau = \frac{1}{c^2}\int_0^\infty \vec{S}\ d\tau = \frac{1}{4\pi}\int_0^\infty\left[\vec{E}\times\vec{H}\right]d\tau , \tag{4.3}$$

where $\Delta\tau$ is the electron field volume, which approximate is equal to $\Delta\tau \cong \zeta\ r_s^3$, where $\zeta$ is a number, and $r_s$ is a dimension of the volume $\Delta\tau$, which contains basic part of the electron energy. Using the quantum form of $U$ and $\vec{S}$:

$$U = \frac{1}{8\pi}\ \psi^+\hat{\alpha}_0\psi , \tag{4.4}$$

$$\vec{S} = -\frac{c}{8\pi}\ \psi^+\hat{\vec{\alpha}}\ \psi = c^2\vec{g} , \tag{4.5}$$

we can analyse the quantum form of (4.1). Since the Dirac's free electron equation solution is the plane wave, we have:

$$\psi = \psi_0\, e^{i(\omega t - ky)}, \qquad (4.6)$$

Taking in to account (4.3) we can write (4.4) and (4.5) in next approximately form

$$\varepsilon_s = U\,\Delta\tau = \frac{\Delta\tau}{8\pi}\,\psi^+\hat{\alpha}_0\psi, \qquad (4.7)$$

$$\vec{p}_s = \vec{g}\,\Delta\tau = \frac{1}{c^2}\vec{S}\,\Delta\tau = \frac{\Delta\tau}{8\pi\,c}\,\psi^+\hat{\vec{\alpha}}\,\psi, \qquad (4.8)$$

Substitute (4.7) and (4.8) into (4.1) and taking in to account (4.4, 4.5), we obtain the following approximate equation:

$$\frac{\partial \psi}{\partial t} - c\hat{\vec{\alpha}}\,\vec{\nabla}\psi + i\frac{\zeta}{2\alpha\,c}\cdot r_s^3\left(\psi^+\hat{\alpha}_0\,\psi - \hat{\vec{\alpha}}\psi^+\hat{\vec{\alpha}}\,\psi\right)\psi = 0, \qquad (4.9)$$

It is not difficult to see that Eq. (4.9) is a non-linear equation of the same type, which was investigated by Heisenberg e.al. [8] and which played for a while the role of the unitary field theory equation. Moreover, Eq. (4.9) is obtained logically rigorously, contrary to the last one. Self-action constant $r_s$ appears in Eq. (4.9) automatically.

## 4.1. Lagrangian of non-linear electromagnetic electron equation

The common form of Lagrangian of the non-linear equation is not difficult to obtain from Lagrangian of the linear Dirac's equation:

$$L_N = \psi^+\left(\hat{\varepsilon} - c\hat{\vec{\alpha}}\,\hat{\vec{p}}\right)\psi + \psi^+\left(\varepsilon_s - c\hat{\vec{\alpha}}\,\vec{p}_s\right)\psi, \qquad (4.10)$$

Using (4.7) and (4.8) we represent (4.10) in explicit quantum form:

$$L_N = i\hbar\left[\frac{\partial}{\partial t}\left[\frac{1}{2}(\psi^+\psi)\right] - c\,div(\psi^+\hat{\vec{\alpha}}\psi)\right] + \frac{\Delta\tau}{8\pi}\left[(\psi^+\psi)^2 - (\psi^+\hat{\vec{\alpha}}\psi)^2\right], \qquad (4.11)$$

Normalising $\psi$-function and using Eqs. (4.2) and (4.3) and transforming Eq.(4.10) in electrodynamics' form we find we obtain:

$$L_N = i\frac{\hbar}{2m_e c^2}\left(\frac{\partial U}{\partial t} + div\,\vec{S}\right) + \frac{\Delta\tau}{m_e c^2}\left(U^2 - c^2\vec{g}^2\right), \qquad (4.12)$$

Here accordingly (3.16) (taking in account that $L=0$) the first summand may be replaced through

$$i\frac{\hbar}{2mc^2}\left(\frac{\partial U}{\partial t}+div\ \vec{S}\right)=\frac{1}{8\pi}\left(\vec{E}^2-\vec{H}^2\right), \quad (4.13)$$

Using the follow known transformation *(which is the electrodynamics' form of Fierz's correlation)* we can now transform the second summand in the follow electromagnetic form:

$$(8\pi)^2\left(U^2-c^2\vec{g}^2\right)=\left(\vec{E}^2+\vec{H}^2\right)^2-4\left(\vec{E}\times\vec{H}\right)^2=\left(\vec{E}^2-\vec{H}^2\right)^2+4\left(\vec{E}\cdot\vec{H}\right)^2, \quad (5.14)$$

So we have:

$$L_N=\frac{1}{8\pi}\left(\vec{E}^2-\vec{H}^2\right)+\frac{\Delta\tau}{(8\pi)^2 m_e c^2}\left[\left(\vec{E}^2-\vec{H}^2\right)^2+4\left(\vec{E}\cdot\vec{H}\right)^2\right], \quad (5.15)$$

As we see, the Lagrangian of the non-linear electron equation contains only the invariant of Maxwell's theory. Let's now analyse the quantum form Lagrangian. The Eq. (4.11) can be written in form:

$$L_N=\psi^+\hat{\alpha}_\mu\partial_\mu\psi+\frac{\Delta\tau}{8\pi}\left[\left(\psi^+\hat{\alpha}_0\psi\right)^2-\left(\psi^+\hat{\vec{\alpha}}\ \psi\right)^2\right], \quad (4.16)$$

Using the quantum form *of Fierz's correlation* (Eq. (4.14))

$$\left(\psi^+\hat{\alpha}_0\psi\right)^2-\left(\psi^+\hat{\vec{\alpha}}\ \psi\right)^2=\left(\psi^+\hat{\alpha}_4\psi\right)^2+\left(\psi^+\hat{\alpha}_5\psi\right)^2, \quad (4.17)$$

from (4.16) we obtain:

$$L_N=\psi^+\hat{\alpha}_\mu\partial_\mu\psi+\frac{\Delta\tau}{8\pi}\left[\left(\psi^+\hat{\alpha}_4\psi\right)^2-\left(\psi^+\hat{\alpha}_5\psi\right)^2\right], \quad (4.18)$$

It is not difficult to see that Lagrangian (4.18) practically coincide with the Nambu's and Jona-Losinio's Lagrangian [9], which is the Lagrangian of the relativistic superconductivity theory.

**Conclusion**

The above results show that the theory of electromagnetic waves can be written in the matrix form also consistently, as in the usual form of Maxwell's theory. Such representation allows looking in a new possibility at connection classical and quantum electrodynamics [10].